\documentclass[a4paper]{article} 

\usepackage[dvipdfmx]{graphicx}
\usepackage[T1]{fontenc}
\usepackage{amssymb}

\newcommand{\argmin}{\mathop{\arg\min}\limits}

\begin{document}

\title{
Refreshing idea on Fourier analysis
}

\author{
Fumihiko Ishiyama
\\
NTT Space Environment and Energy Labs.\\
Nippon Telegraph and Telephone Corp.\\
Tokyo, Japan \\
fumihiko.ishiyama@ntt.com
}

\date{}

\maketitle

\begin{abstract}
The  
``theoretical limit of time-frequency resolution in Fourier analysis''
is thought to originate in certain mathematical and/or physical limitations.
This, however, 
is not true.
The actual origin arises from the 
numerical (technical) method deployed to reduce computation time.
In addition, there is a gap between the theoretical equation
for Fourier analysis  
and its numerical implementation.
Knowing the facts 
brings us practical benefits.
In this case, these related to
boundary conditions, and complex integrals.
For example, 
replacing a Fourier integral with a complex integral brings
a hybrid method
for the 
Laplace and Fourier transforms,
and reveals another perspective on time-frequency analysis.
We present such a perspective here with a simple demonstrative analysis.
\end{abstract}


\renewcommand{\thefootnote}{\fnsymbol{footnote}}
\footnote[0]
{
\copyright 2025 IEEE. 
Personal use of this material is permitted. 
Permission from IEEE must be obtained for all other uses, 
in any current or future media, 
including reprinting/republishing this material 
for advertising or promotional purposes,
creating new collective works, for resale or redistribution
to servers or lists, or reuse of any copyrighted component
of this work in other works.
}
\renewcommand{\thefootnote}{\arabic{footnote}}

\section{Introduction}

The 
``theoretical limit of time-frequency resolution of Fourier analysis''
is thought to originate in certain mathematical and/or physical limitations,
such as
``it is from quantum mechanics $|\Delta f \Delta t| \geq h/2 \pi$,''
where $h$ is Planck constant.

This, 
however, is not true.
The actual origin arises from the numerical method deployed to reduce computation time.
In addition, there is a gap between the theoretical equation for Fourier analysis 
and its numerical implementation.

Let us remind ourselves of the equation for Fourier transform
\begin{equation}\label{fourier}
S(f)=\int_{-\infty}^\infty s(t) e^{- 2 \pi i f t} dt.
\end{equation}
The equation requires an infinite continuous time series.
However, the actual time series which in practice we have for analysis
is always finite and discrete, 
and does not satisfy this requirement of the equation.
Therefore, some modifications 
have to be made to 
 the finite and discrete time series in order
to satisfy the requirements of the equation,
and 
it is these
modifications
that are the origin of the time-frequency limitation
currently under discussion.

For this reason, we are using this article to
 present a refreshing idea.
Breaking the time-frequency resolution limitation is a practically useful proposition.

\begin{figure}[htb]
  \centerline{\includegraphics[width=90mm]{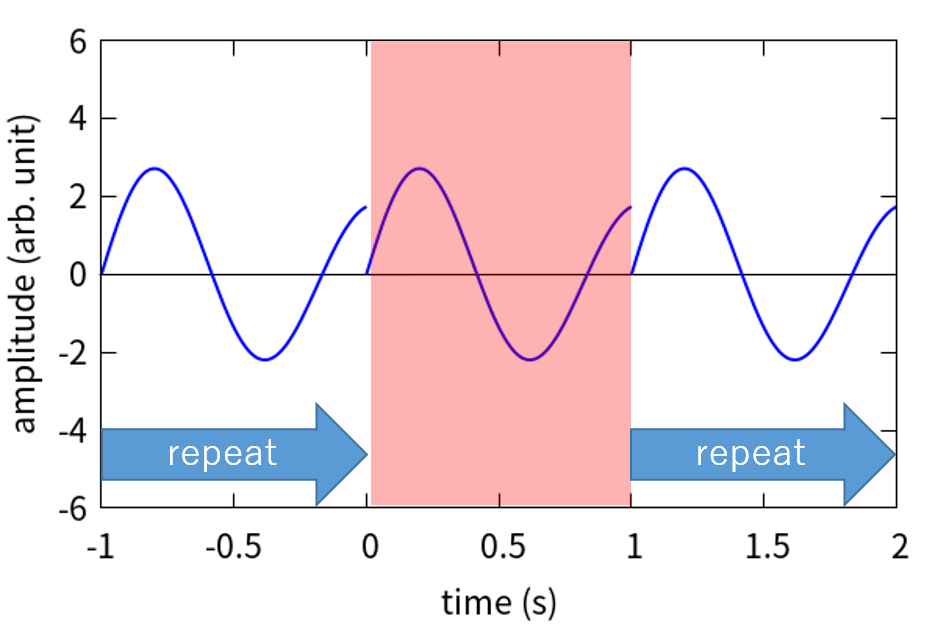}}
\caption{Periodic boundary condition.}
\label{fig-pbc}
\end{figure}

First, we need an infinite time series.
For this purpose, the ``periodic boundary condition''
which is shown in Fig.~\ref{fig-pbc}, is commonly introduced~\cite{walker}.
This condition is implicitly applied to all conventional linear methods,
and replacing it with an alternative condition is the first step in the process of  
clearing  the time-frequency resolution limitation.
This means that all of the conventional linear methods share limitation in common
that,
as we will demonstrate below, can be overcome.

What the periodic boundary condition does is as follows.
Assume that the finite time series, which is  available for analysis,
lines  in the hatched area of the figure.
From this position, 
we simply repeat the hatched time series infinitely, to fill the infinite time series.

Now, we have an infinite time series for the Fourier transform.
However, 
the preparatory step shown in Fig.~\ref{fig-pbc} throws up some awkward questions.
It is undeniably the case that
the possible frequencies are limited to the harmonics within the hatched area.
Even though we have an infinite time series, the harmonics are limited,
and herein lies the origin of the 
time-frequency resolution limitation shared by all the conventional linear methods.
Indeed, this condition does boast the merit of reducing 
 computation time~\cite{walker,ccisp},
and has been historically useful. 
However we would suggest that this merit is an anachronism in any case, 
on account of the improvements we have witnessed in computation power.

\begin{figure}[htb]
  \centerline{\includegraphics[width=90mm]{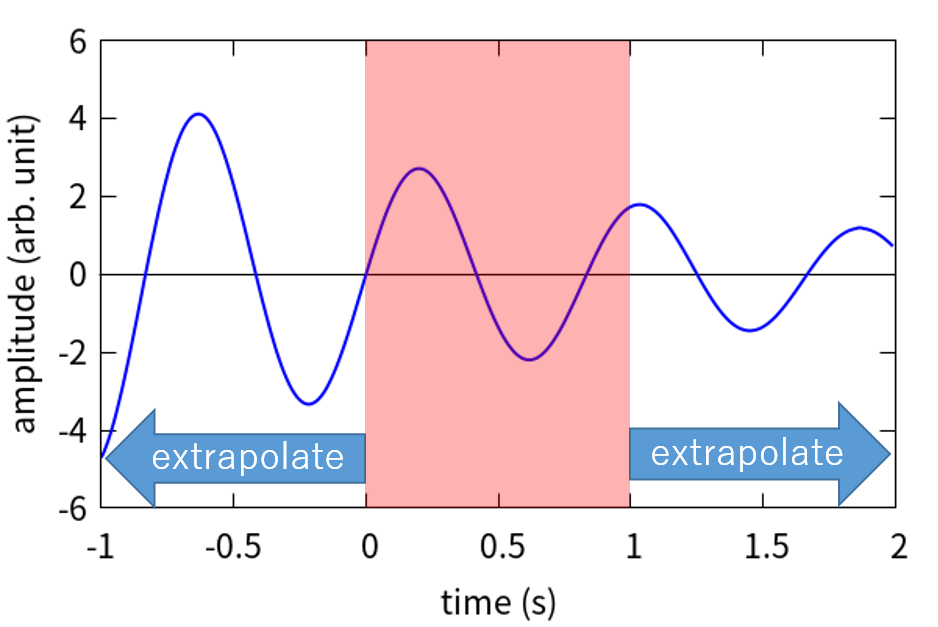}}
\caption{Proposed linear extrapolation condition.}
\label{fig-lxc}
\end{figure}

Therefore, we suggest replacing  this condition with an alternative
to fill out the infinite time series, and to clear the time-frequency resolution limitations
that the conventionally applied condition introduces.

Our choice for an alternative is the linear extrapolation condition, shown in Fig.~\ref{fig-lxc}.
However, this condition itself is unbound for $t \rightarrow \pm \infty$, 
and some tricks are required~\cite{nolta2023}.

\newpage

In the following sections, 
we expand the Fourier analysis with some tricks, 
present a demonstrative analysis with the use of frequency modulated time series, 
and conclude this paper.

\section{Expanding Fourier analysis}

Our method looks very different from a conventional Fourier analysis, 
because of various unfamiliar concepts in the model equation.
However, 
it will be found that the proposed method is a natural expansion of 
the familiar Fourier analysis.

Our method of analysis is not based on a Fourier integral, 
but on a mode decomposition 
with general complex functions, 
which organize nonlinear oscillators \cite{nolta2023},
from which  we calculate the local linearized solution \cite{kubo} of the decomposition.

More complete 
details of our method are set out
in previously published papers~\cite{ccisp,nolta2023,cspa,isspit},
and we ask the readers to please refer to them for further understanding.
What we present here is  just an outline.

\subsection{Model equation}

We expand the given time series $S(t) \in \mathbb{R}$ with 
general complex functions $H_m(t) \in \mathbb{C}$ as 
\begin{equation}\label{eq-noe}
S(t) = \sum_{m=1}^M e^{H_m(t)},
\end{equation}
where $M$ is the number of complex functions.

The complex functions are expressed as 
\begin{equation}\label{eq-h-fl}
H_m(t) = \ln c_m(t_0) + \int_{t_0}^t [ 2 \pi i f_m(\tau) + \lambda_m(\tau) ] {\rm d}\tau,
\end{equation}
where $f_m(t)~\in~\mathbb{R}$ represents the frequency modulation (FM) terms,
which is known as instantaneous frequency~\cite{pol} by van der Pol,
$\lambda_m(t)~\in~\mathbb{R}$ represents the amplitude modulation (AM) terms,
which is our original~\cite{nolta2023,cqg},
and $c_m(t_0) \in \mathbb{C}$ represents the amplitudes of the oscillators
at $t=t_0$.

This expansion corresponds to a mode decomposition with general complex functions,
noting that 
\begin{equation}
H_m^\prime(t) = 2 \pi i f_m(t) + \lambda_m(t).
\end{equation}

Additionally, note that
the case
\begin{equation}
H_m^\prime(t) = 2 \pi i \frac{m}{M \Delta T},
\end{equation}
becomes a Fourier series expansion
\begin{equation}
S(t) = \sum_{m=1}^M e^{H_m(t)}=\sum_{m=1}^M c_m(t_0) e^{2 \pi i \frac{m}{M \Delta T} (t-t_0)},
\end{equation}
itself.
That is, our model equation contains Fourier analysis 
as a special case,
and is a natural expansion of the analysis.

\subsection{Locally linearized solution}

As it is known that our model equation does not have a unique solution, 
due to the non-linearity \cite{daubeshies2011,cspa2023},
we need a special idea to make our model equation uniquely solvable~\cite{ccisp,kubo}.
This is a part of our tricks, which corresponds to the linear extrapolation condition
shown in Fig.~\ref{fig-lxc}.

We expand our model equation Eq.~(\ref{eq-noe}) as 
\begin{equation}\label{eq-exp-h-local}
S(t)|_{t \sim t_k} \simeq \sum_{m=1}^M e^{H_m(t_k)+H_m^\prime(t_k) (t-t_k) + 
O\left((t-t_k \right)^2)},
\end{equation}
around $t \sim t_k=t_0+k \Delta T$,
consider a short enough time width, 
and ignore the higher order terms $O((t-t_k)^2)$.

Then, 
the equation becomes a simple linear equation
\begin{eqnarray}
\label{eq-exp-h-locallinear}
S(t)|_{t \sim t_k} &\simeq& \sum_{m=1}^M e^{H_m(t_k)+H_m^\prime(t_k) (t-t_k) }
\\
\label{eq-exp-h-locallinear-b}
&=& \sum_{m=1}^M c_m(t_k) e^{[2 \pi i f_m(t_k)+\lambda_m(t_k)] (t-t_k) },
\end{eqnarray}
and we can obtain unique $H_m^\prime(t_k)$ easily 
by applying the numerical method of  
linear predictive coding (LPC) with $N$ samples, 
noting that we must use a non-standard numerical method~\cite{ccisp,git}
to hold the condition shown in Fig.~\ref{fig-lxc}.

The standard method of LPC is not adequate,
because it contains
the unfavorable condition  \cite{walker} shown in Fig.~\ref{fig-pbc}, 
and
an approximation \cite{itakura} to reduce 
computation cost.

Subsequently, we calculate 
the complex amplitudes $c_m(t_k)$
of the oscillators
$c_m(t_k) e^{H_m^\prime(t_k) (t-t_k)}$ as
\begin{equation}\label{eq-calc-amps}
\argmin_{c_{m}(t_k)}
\sum_{n=0}^{N-1} 
\left(
S(t_k+n\Delta T)
-
\sum_{m=1}^M c_m(t_k) e^{n H_m^\prime(t_k) \Delta T}
\right)^2.
\end{equation}

\subsection{Instantaneous spectrum}

The equation for instantaneous spectrum is given as follows.

We transform the locally linearized time series  $S(t)|_{t \sim t_k}$ around $t \sim t_k$
(Eq.~(\ref{eq-exp-h-locallinear-b}))
with a complex integral
as
\begin{eqnarray}\label{eq-specg-cont}
F(f,t_k)&=& \int_C S(t)|_{t \sim t_k} e^{-2 \pi i ft} dt 
\nonumber
\\
&=& 
\sum_m \int_C c_m(t_k) e^{[\lambda_m(t_k)+ 2 \pi i f_m(t_k)]t -2 \pi i ft} dt 
\nonumber
\\
&=& 
\sum_m c_m(t_k) \int_C  e^{[\lambda_m(t_k)+ 2 \pi i (f_m(t_k)-f)]t} dt 
\nonumber
\\
&=& 
\sum_m \frac{ c_m(t_k) }{\lambda_m(t_k)+ 2 \pi i (f_m(t_k)-f)}.
\end{eqnarray}

Note that the Laplace transform is written as
\begin{equation}
L(s)=\int_0^\infty s(t) e^{-st} dt =i \int_0^{-i \infty} s(i\tau) e^{- i s\tau} d\tau,
\end{equation}
and is understood as a Fourier integral on the imaginary axis.
Therefore, the Laplace transform of $S(t)|_{t \sim t_k}$ around $t \sim t_k$ becomes
\begin{equation}
\Lambda(\lambda, t_k)=\sum_m \frac{ c_m(t_k) }{(\lambda_m(t_k)-\lambda)+ 2 \pi i f_m(t_k)}.
\end{equation}

As Eq.~(\ref{eq-specg-cont}) is for a continuous time series, 
some modifications for its application to a discrete time series are required.
That is, specifically, a modification from a Fourier transform to a Fourier series expansion.

For example, when $\lambda_m(t_k) = 0$, 
the equation is unbound at $f=f_m(t_k)$, and is not practical.
The practical value is the maximum value $|c_m(t_k)|$ for discrete system.

Therefore, we take the absolute value of each term, and 
adjust the maximum values at  $f=f_m(t_k)$ so as to be $|c_m(t_k)|$ \cite{ccisp}.

\begin{equation}\label{eq-specg-disc}
F_{\rm disc}(f,t_k)=\sum_m 
\left| \frac{ c_m(t_k) \lambda_m(t_k) }{\lambda_m(t_k)+ 2 \pi i (f_m(t_k)-f)} \right|
\end{equation}

This equation has several merits~\cite{ccisp,nolta2023}. For example,
the instantaneous spectrum of each term is available,
and this feature is valuable for signal separation,
as we show below.
In addition, 
$F_{\rm disc}(f,t_k)^2$ becomes power spectrum, 
corresponding to the conventional Fourier power spectrum.

\subsection{Revised time-frequency resolution} 

We briefly demonstrate 
how our method works
\cite{git}.
The source code and its execution output of the followings are shown in the reference.

\begin{figure}[htb]
  \centerline{\includegraphics[width=90mm]{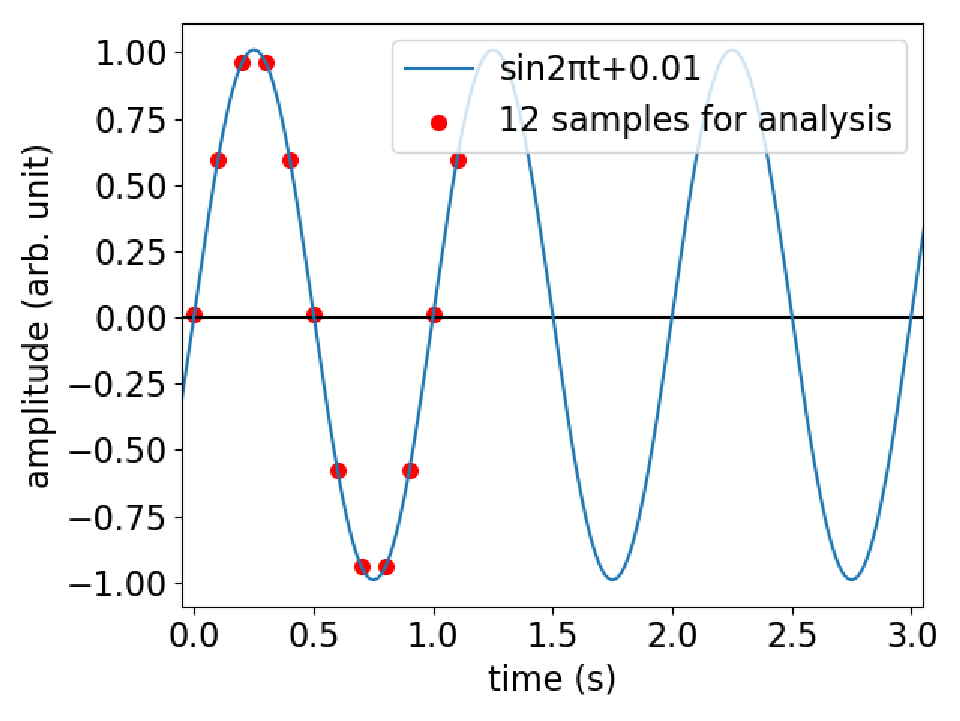}}
\caption{
Time series for analysis. 
}
\label{fig-git-ts}
\end{figure}

The time series for analysis shown in Fig.~\ref{fig-git-ts} is
\begin{equation}
S(t)= 0.01 + \sin 2 \pi t,
\end{equation}
and we take twelve samples
with a sampling frequency of $10$~Hz,
as shown in the figure.
The samples correspond to $1.2$ cycles of the oscillation,
and 
this small sample set and short time series are sufficient for our method \cite{cspa2023},
in contrast to conventional methods.

\begin{figure}[htb]
  \centerline{\includegraphics[width=90mm]{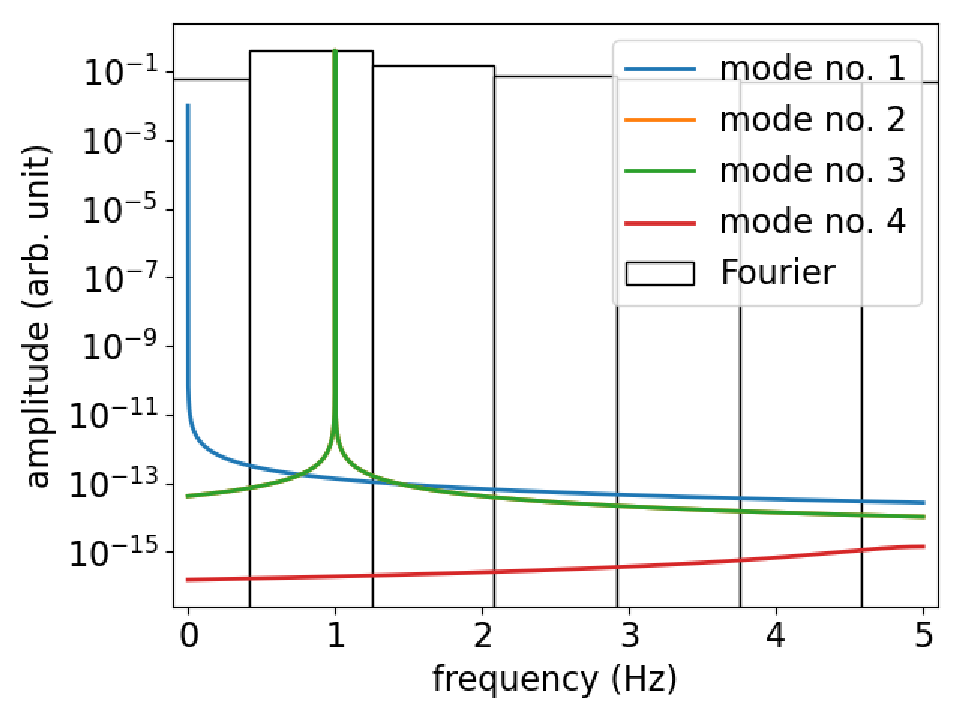}}
\caption{
Spectrum of each obtained mode.
}
\label{fig-git-spec}
\end{figure}

Following this, we apply our method to the twelve samples, 
with the parameters $M=7,~N=12$,
and 
plot each obtained term in Eq.~(\ref{eq-specg-disc})
in Fig.~\ref{fig-git-spec}.

In addition, we plot the conventional Fourier spectrum from the same twelve samples
in the figure,
which corresponds to a bin of short time Fourier transform (STFT).

Four spectra 
(no.~1 to 4),
are shown in the figure.
Each spectrum corresponds to 
(no.~1)
a constant term with amplitude $0.01$,
 (no.~2 and 3, the same spectra)
a sinusoidal time series with a frequency of $1$ Hz,
and
 (no.~4)
computational error, 
which corresponds to white noise on the time series.

Note that white noise on the given time series (no.~4) is obtained 
as a mode with a flat spectrum,
and we can remove this unnecessary term
from Eq.~(\ref{eq-specg-disc}) for plotting 
spectrum.

Obtained numerical results for no. 1 to 3 are shown in Table~\ref{tab-tf-resolution}.

\begin{table}[htb]
\caption{numerical resolutions}
\label{tab-tf-resolution}
\centerline
{
\begin{tabular}{c|c|c}
term & frequency (Hz) & amplitude
\\
\hline 
no. 1 &
0.0~~~~~~~~~~~~~~~~~~~~~ &
0.009999999999492205
\\
\hline 
no. 2 \& 3 &
~0.9999999999999438 & 
~1.0000000000004858~~~~
\\
\hline
\end{tabular}
}
\end{table}

Conventional Fourier analysis 
requires the time width $T$~s to obtain the frequency resolution $f_{\rm res}=1/T$~Hz,
and it holds that $T \times f_{\rm res}=1$.
This is the limit of the time-frequency resolution for a  conventional Fourier analysis.

In contrast, 
the frequency resolution $f_{\rm res}$ of no.~2\&3 in Table~\ref{tab-tf-resolution}
is $1-0.9999999999999438=5.62\cdot10^{-14}$~Hz,
and is obtained using the time width $T=1.2$~s.
Therefore, the time-frequency resolution
becomes 
$T \times f_{\rm res}=6.74~\cdot~10^{-14}$, 
and this resolution corresponds to the computational resolution
of the numerical data.
That is, the time-frequency resolution of our method is bounded by 
the resolution of the given numerical data,
and not by the method itself.

\section{Demonstrative analysis}

Now we show a sample of a spectrogram using Eq.~(\ref{eq-specg-disc}).
For the purpose, we employ a simple time series $S(t)$ with frequency modulation
\begin{equation}
S(t)
=
\sin 2 \pi \left( ft+a\cos w t \right).
\label{eq-fm-ts}
\end{equation}

Assuming the parameters in  Eq.~(\ref{eq-fm-ts}) as
$f=100$~Hz, $a=0.5$, and $w=10$,
the time series of modulated frequency $f(t)$ becomes
\begin{eqnarray}
f(t) 
&=&
\left( ft+a\cos w t \right)^\prime
\nonumber
\\
&=& f - a w \sin w t 
\nonumber
\\
&=& 100-5 \sin 2 \pi \frac{10}{2 \pi} t
\nonumber
\\
&\simeq& 100-5 \sin 2 \pi \frac{1}{0.63} t.
\label{eq-fm-ts-f}
\end{eqnarray}

Note 
that the waveform $-\sin w t$ appears in the time series.

\begin{figure}[htb]
\centerline{\includegraphics[width=90mm]{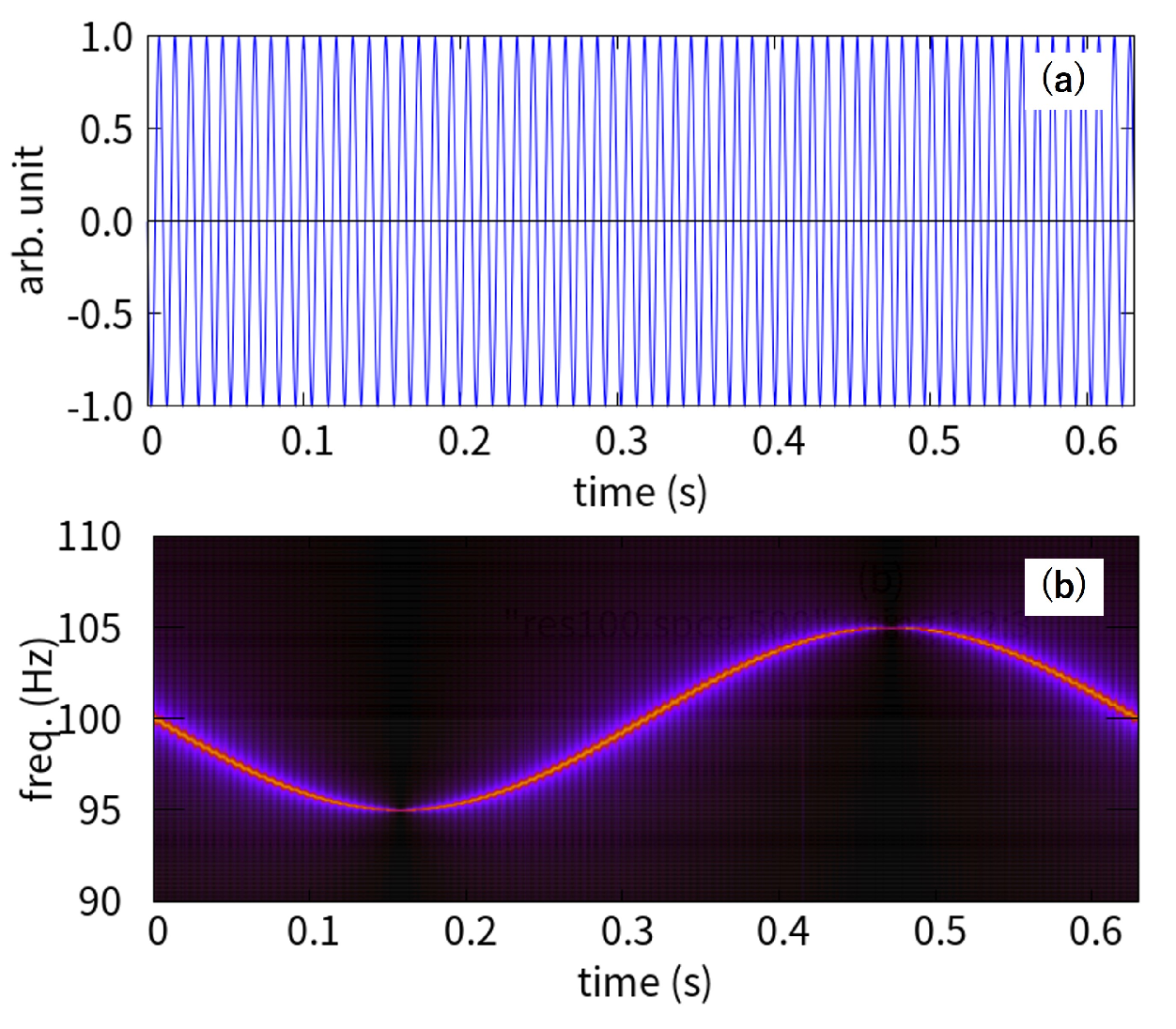}}
\caption{
(a) Time series for analysis Eq.~(\ref{eq-fm-ts}), 
and 
(b) obtained spectrogram Eq.~(\ref{eq-fm-ts-f}).
Area of interest is magnified.
}
\label{fig-fm}
\end{figure}

We show the time series for analysis Eq.~(\ref{eq-fm-ts}) and 
its corresponding spectrogram Eq.~(\ref{eq-specg-disc}), 
which has the time series of modulated frequency in Eq.~(\ref{eq-fm-ts-f}),
in Fig.~\ref{fig-fm}.

The sampling frequency $f_s$ was set to $10$~kHz, 
and 
the other parameters for the analysis were set to
$M=10$, and $N=20$.

Therefore, 
the range of the spectrogram is $5$~kHz, and the enlarged view 
(20~Hz width) is shown in the figure.
In addition, the time width for a single analysis becomes $N f /f_s=0.2$ cycles of 
the base frequency
$f=100$~Hz. 
This time width is sufficient for our method of analysis, as is shown in Fig.~\ref{fig-fm}.

The upshot of all this is 
that a ``less-than-a-cycle'' time-frequency analysis is available 
with our method of analysis, 
with the resolution shown in Fig.~\ref{fig-fm} 
and also  in Table~\ref{tab-tf-resolution}.

\section{Conclusion}

We presented a refreshing idea on Fourier analysis.
The idea makes less-than-a-cycle time-frequency analysis a reality,
offers signal to noise manipulation, 
and brings mathematical understanding on
nonlinear systems, such as the origin of the modulated frequencies.

We think that our method will not be the final nor ultimate way of time-frequency analysis,
but we believe it to be sufficiently  effective in its current form.

The know-how required to apply
 our method,
 such as the setting of parameters,
  is still under investigation,
and requires further work.


\end{document}